\documentclass[preprint,showpacs,preprintnumbers,amsmath,amssymb]{revtex4}

\usepackage{graphicx}
\usepackage{dcolumn}
\usepackage{bm}

\begin{document}


\title{Crystal structure, Fermi surface calculations and \\ Shubnikov-de Haas oscillations spectrum of the organic
metal $\theta$-(BETS)$_4$HgBr$_4$(C$_6$H$_5$Cl) at low
temperature}

\author{ David Vignolles$^{1}$, Alain Audouard$^{1\dagger}$, Rustem B. Lyubovskii$^{2}$,\\ Sergei I. Pesotskii$^{2}$,
J\'{e}r\^{o}me B\'{e}ard$^{1}$, Enric Canadell$^{3}$, Gena V.
Shilov$^{2}$, Olga A. Bogdanova$^{2}$, Elena I. Zhilayeva$^{2}$
and Rimma N. Lyubovskaya$^{2}$}

\affiliation{$^1$ Laboratoire National des Champs Magn\'{e}tiques
Puls\'{e}s (UMS CNRS-UPS-INSA 5147), 143 avenue de Rangueil, 31400
Toulouse, France. \\$^2$Institute of Problems of Chemical Physics,
Russian Academy of Sciences, 142432 Chernogolovka, MD,
Russia\\$^3$Institut de Ci\`{e}ncia de Materials de Barcelona,
Consejo Superior de Investigationes Cient\'{i}ficas, Campus
Universitat Aut\`{o}noma de Barcelona,
Bellaterra 08193, Spain}%

\date{\today}

\begin{abstract}
The organic metal $\theta$-(BETS)$_4$HgBr$_4$(C$_6$H$_5$Cl) is
known to undergo a phase transition as the temperature is lowered
down to about 240 K. X-ray data obtained at 200 K indicate a
corresponding modification of the crystal structure, the symmetry
of which is lowered from quadratic to monoclinic. In addition, two
different types of cation layers are observed in the unit cell.
The Fermi surface (FS), which can be regarded as a network of
compensated electron and hole orbits according to band structure
calculations at room temperature, turns to a set of two
alternating linear chains of orbits at low temperature. The field
and temperature dependence of the Shubnikov-de Haas oscillations
spectrum have been studied up to 54 T. Eight frequencies are
observed which, in any case, points to a FS much more complex than
predicted by band structure calculations at room temperature, even
though some of the observed Fourier components might be ascribed
to magnetic breakdown or frequency mixing. The obtained spectrum
could result from either an interaction between the FS's linked to
each of the two cation layers or to an eventual additional phase
transition in the temperature range below 200 K.
\end{abstract}



\maketitle $\dagger$ Electronic address: audouard@lncmp.org

\section{\label{sec:intro}Introduction}
In high enough magnetic fields, the Fermi surface (FS) of many
quasi-two-dimensional (q-2D) organic metals is liable to give rise
to networks of orbits coupled by magnetic breakdown (MB)
\cite{Ka04}. Magnetic oscillations spectra of such compounds often
contain Fourier components of which the frequencies are linear
combinations of few basic frequencies. Some of these components
correspond to MB-induced closed orbits that are accounted for by
the semi-classical model of Falicov and Stachowiak \cite{Fa66}. In
the case of Shubnikov-de Haas (SdH) oscillations, quantum
interference (QI) paths \cite{St71} have also been invoked.
However, the so called "forbidden frequencies" observed in the de
Haas-van Alphen  spectra of organic metals that illustrate the
linear chain of coupled orbits model by Pippard \cite{Pi62} cannot
be interpreted on these bases. Indeed, they are rather attributed
to frequency mixing due to the oscillation of the chemical
potential \cite{mu} and (or) to the MB-induced modulation of the
density of states \cite{Pi62,LLB} even though the respective
influence of these two contributions remains to be determined.

Another type of network is provided by the family of organic
metals (BEDT-TTF)$_8$Hg$_4$Cl$_{12}$(C$_6$H$_5$X)$_2$ (X = Cl, Br)
whose FS, which originates from two pairs of crossing q-1D sheets,
is composed of one electron and one hole tube with the same area
\cite{Ve94}. As it is the case of the above-mentioned linear
chains of coupled orbits, SdH oscillations spectra in this type of
network also reveal frequency mixing \cite{Pr02, Vi03, Au05}.
However, in striking contrast to the data relevant to linear
chains of orbits, all the Fourier components observed in the de
Haas-van Alphen oscillations spectra can be consistently
interpreted on the basis of the model of Falicov and Stachowiak
\cite{Au05}. It is thus of primary importance, from the viewpoint
of the quantum oscillation physics, to check whether or not the
above mentioned behaviour is observed in other organic metals
whose FS can be regarded as networks of compensated orbits. Such
networks can also be realized in organic metals of which the FS
originates from the overlapping in two (or more) directions of
hole tubes with an area equal to that of the First Brillouin zone
(FBZ) and from the resulting gap openings \cite{Ro04}. According
to band structure calculations, this is the case of organic metals
such as (BEDO-TTF)$_4$ReO$_4$$\cdot$H$_2$O \cite{Kh98} and $\beta
^{\prime\prime}$-(BEDT-TTF)$_4$(NH$_4$)[M(C$_2$O$_4$)$_3$]$\cdot$DMF
(M = Fe, Cr) \cite{Pr03}. However, SdH oscillations spectra of
these compounds did not yield frequency combinations in a field
range large enough in order to perform a reliable analysis of the
relevant data \cite{Pr01,Au04,Au06}.

\begin{figure} [h]                                                    
\centering
\resizebox{\columnwidth}{!}{\includegraphics*{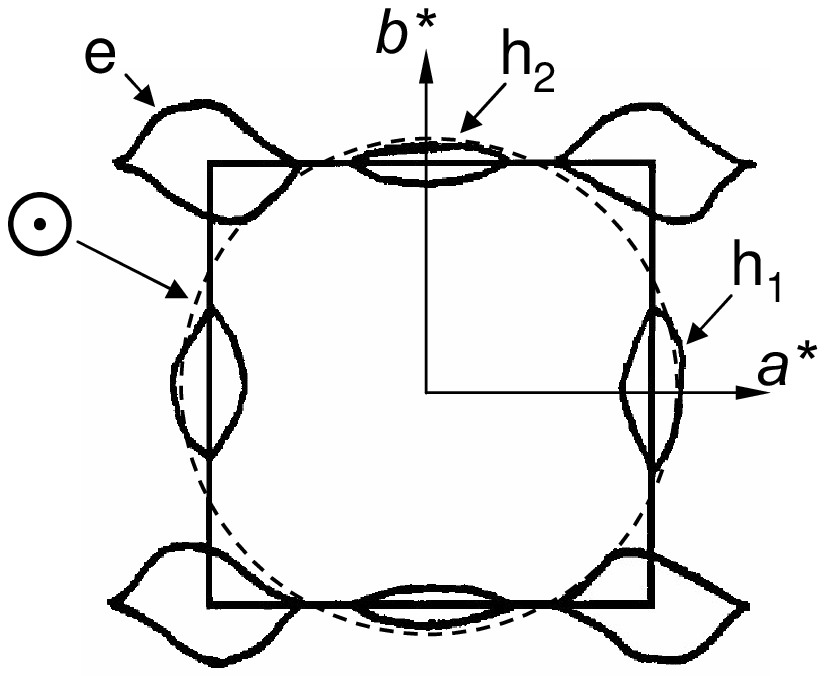}}
\caption{\label{Fig:FS_roomT} Fermi surface (FS) for a single
donor layer of $\theta$-(BETS)$_4$HgBr$_4$(C$_6$H$_5$Cl) at room
temperature according to band structure calculations of Ref.
\cite{Ly01}. The dashed line corresponds to the $\bigodot$ orbit,
from which the FS is built, with an area equal to that of the
First Brillouin zone (see text). Since the tetragonal cell
contains four identical donor layers rotated by 90 degrees from
each other around the c-axis, the resulting FS of the system could
be more complex.}
\end{figure}

According to band structure calculations based on room temperature
X-ray data \cite{Ly01}, the FS of the q-2D organic metal
$\theta$-(BETS)$_4$HgBr$_4$(C$_6$H$_5$Cl) is composed of two hole
and one electron tubes (see Fig. \ref{Fig:FS_roomT}). Since the
electron- and hole-type tubes are compensated, the FS should yield
the same kind of network as above discussed. However, a phase
transition is observed around 240 K which is liable to modify the
FS. As a matter of fact, the SdH oscillations spectra recorded in
the field range below 15 T only exhibit two frequencies at 40 T
and 210 T \cite{Ly01} which is not in agreement with the FS
calculations at room temperature. The aim of this article is
therefore twofold. In a first step, crystal and electronic band
structures based on X-ray data collected at a temperature below
the phase transition, namely 200 K, are reported and, in a second
step, the SdH oscillations spectrum is explored in the high field
range in order to reveal other possible frequencies.

\section{Experimental}                                                
The crystals studied were synthesized by the
electrocrystallization technique reported in Ref. \cite{Ly01}.
Their approximate dimensions were 0.4 $\times$ 0.15 $\times$ 0.1
mm$^3$ for X-ray diffraction experiments and 1.5 $\times$ 1.0 $\times$ 0.1 mm$^3$
for  magnetoresistance measurements.

Diffraction measurements were performed at a temperature of 200 K
with a P4 Bruker AXS diffractometer ($\lambda$(Mo K$\alpha$) =
0.71073 $\AA$, $\theta$/2$\theta$ scanning). The main
crystallographic data are:
(C$_{10}$H$_8$S$_4$Se$_4$)$_4$HgBr$_4$-C$_6$H$_5$Cl, monoclinic,
space group Cc, a = 13.678(3) $\AA$, b = 75.857(16) $\AA$, c =
9.533(2) $\AA$, V = 7128(3) $\AA^3$, $\beta$ = 133.89$^{\circ}$, Z
= 4, d$_{cal}$ = 2.723 g cm$^{-3}$, F(000) = 5400. The unit cell
parameters were determined and refined using 35 reflections in the
range 5$^{\circ}  <$ $\theta <$  20$^{\circ}$. Within a set of
6531 X-ray reflections taken in the interval 0$^{\circ}  < \theta
<$ 50$^{\circ}$, 4213 crystallographically independent ones had
intensities I$>$ 2$\sigma$(I). The structure was solved by a
direct method. The positions of non-carbon atoms were refined in
the anisotropic approximation by the full-matrix least square
method. The positions of carbon atoms were refined in the
isotropic approximation only. The positions of the atoms belonging
to the C$_6$H$_5$Cl solvent molecule were revealed from difference
Fourier syntheses and refined within the isotropic approximation
with restrictions imposed on the bond lengths. The positions of
the hydrogen atoms cannot be revealed by the difference Fourier
syntheses. The structure was refined from the independent
reflections with I $>$ 2$\sigma$(I), using the SHELXL-97 program
package to R1 = 0.1235. It should be noticed that as the
temperature decreases the above mentioned phase transition yields,
at a macroscopic level, lower quality single crystals. More
specifically, the X-ray peaks broadened, some of them split, i.e.
a single crystal become twinned, that prevents reliable X-ray
experiments. For one of the slowly cooled single crystals studied,
the contribution of one of the twin components was higher than
that of the other one after the phase transition. It should be
noted that this crystal was far from being an ideal one which
justifies the poor R1 value above mentioned. The obtained
experimental massif of reflections was somewhat distorted because
of the low quality that was manifested in the structure refinement
parameters and did not allow for a high resolution of the
molecular structure because of significant uncertainties in the
determination of interatomic distances and angles. Nevertheless,
we were able to perform the X-ray experiments in corpore. In
addition, as developed in Section \ref{sec:SdH}, clear quantum
oscillations are observed.

The tight-binding band structure calculations were based upon the
effective one-electron Hamiltonian of the extended H\"{u}ckel
method \cite{Wh78}. The off-diagonal matrix elements of the
Hamiltonian were calculated according to the modified
Wolfsberg-Helmholz formula \cite{Am78}. All valence electrons were
explicitly taken into account in the calculations and the basis
set consisted of double-$\zeta$ Slater-type orbitals for C, S and
Se and single-$\zeta$ Slater-type orbitals for H. The exponents,
contraction coefficients and atomic parameters for C, S, Se and H
were taken from previous works \cite{Pe90, Pe93}.

Two crystals, labelled  $\# 1$ and $\# 2$ in the following, were
studied in pulsed magnetic fields in the temperature range from
1.6 K to 4.2 K. The maximum field and pulse decay duration were 54
T, 0.32 s and 36 T, 1 s for crystal $\# 1$ and $\# 2$,
respectively. Electrical contacts to the crystals were made using
annealed platinum wires of 20 $\mu$m in diameter glued with
graphite paste. A one-axis rotating sample holder allowed to
change the direction of the magnetic field with respect to the
crystallographic axes for crystal $\# 2$. Alternating current
(1$\mu$A, 77 Hz and 5 $\mu$A, 20-50 kHz for zero-field and
magnetoresistance measurements, respectively) was injected
parallel to the \emph{b} direction (interlayer configuration). A
lock-in amplifier with a time constant of 100 ms and 100 $\mu$s
for zero-field resistance and magnetoresistance measurements,
respectively, was used to detect the signal across the potential
contacts. In the following, the absolute value of the amplitude of
the magnetoresistance oscillations is obtained through discrete
Fourier transforms calculated with a Blackman window.

\section{\label{sec:results}Results and discussion}

\subsection{\label{sec:structures}Crystal and electronic band
structures}                                                           

\begin{figure}                                                    
\centering
\resizebox{\columnwidth}{!}{\includegraphics*{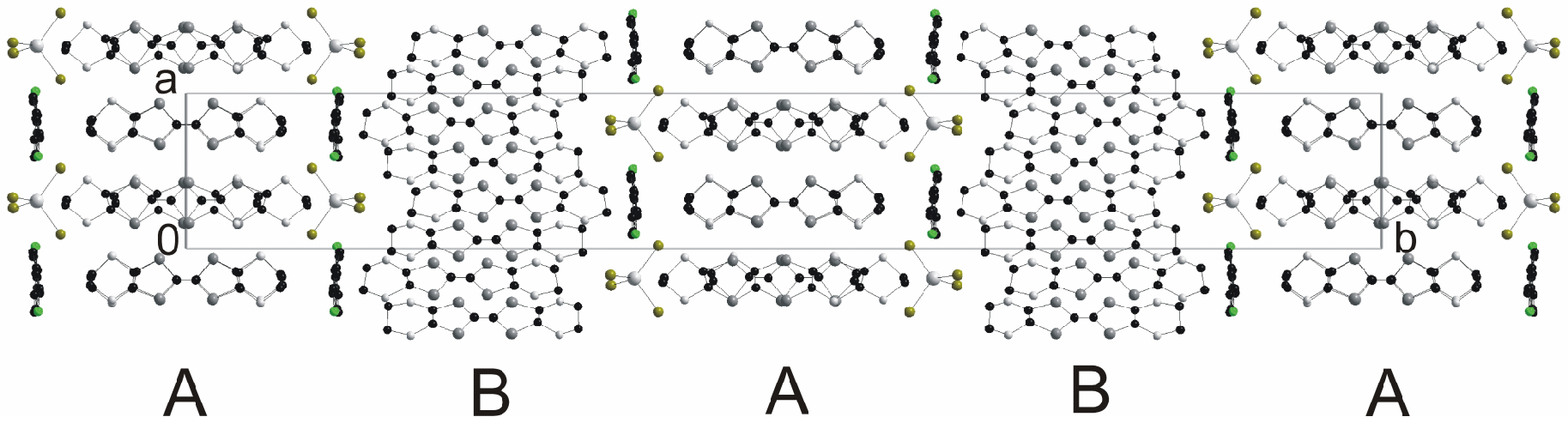}}
\caption{\label{Fig:ab_plane} Crystal structure of
$\theta$-(BETS)$_4$HgBr$_4$(C$_6$H$_5$Cl) projected along the
\emph{c} axis. The labels A and B refer to the two different
cation layers (see text).}
\end{figure}

\begin{figure}                                                    
\centering
\resizebox{\columnwidth}{!}{\includegraphics*{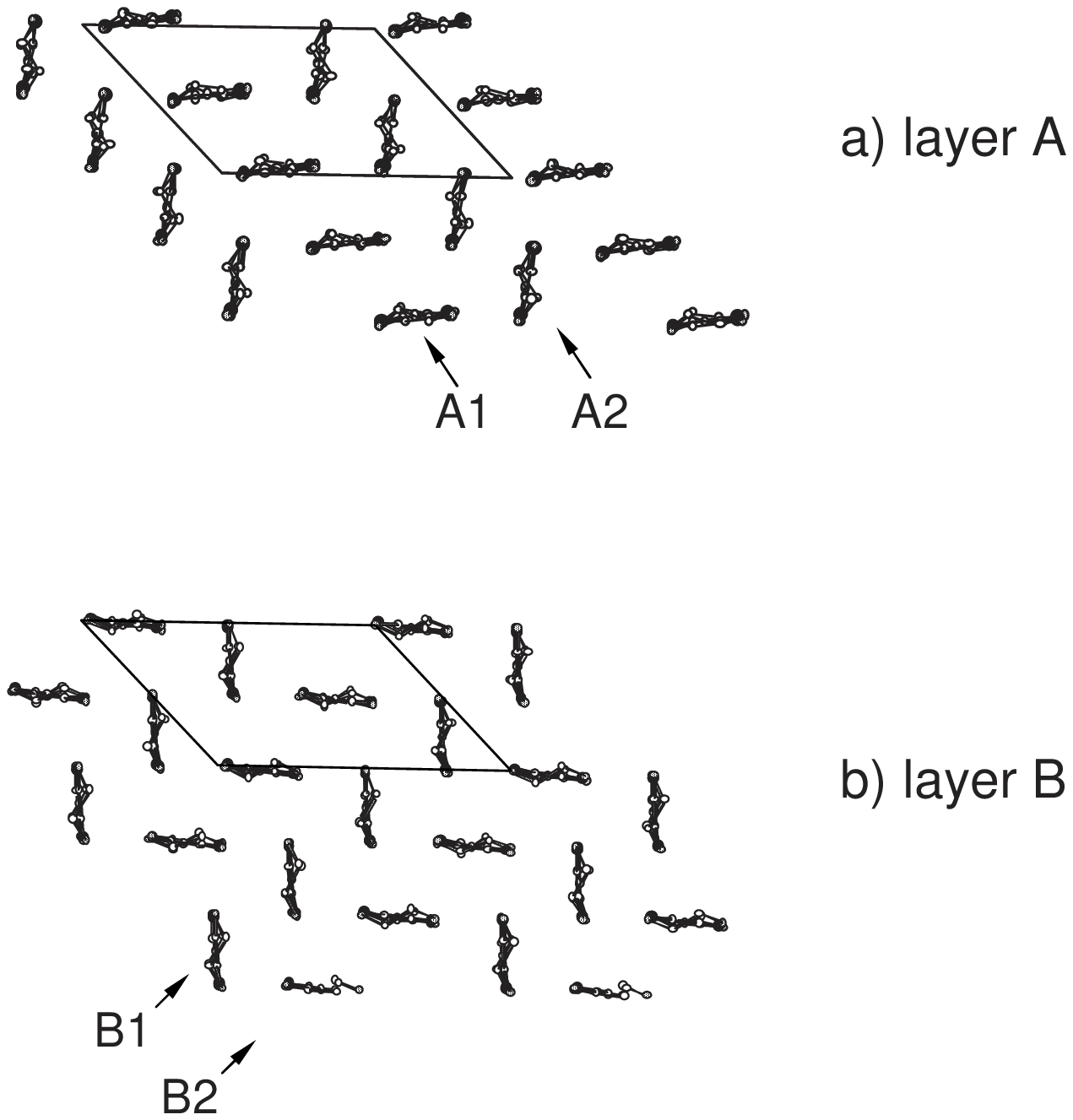}}
\caption{\label{Fig:layer_A_B} Conducting cation layers A and B.}
\end{figure}

The independent part of the crystal structure involves four BETS
molecules, one HgBr$_4^{2-}$ ion  and one solvent molecule which
can occupy two sites with a probability of 50$\%$ for each of
them. Fig. \ref{Fig:ab_plane} displays the projection of the
crystal structure on the \emph{ab} plane. As it is the case at
room temperature, the unit cell contains four radical cation
layers, between which are located the HgBr$_4^{2-}$ ions and the
solvent (C$_6$H$_5$Cl) molecules. In contrast with the room
temperature data \cite{Ly01}, two different kinds of cation
layers, referred to as A and B in the following, are evidenced at
200 K (see Figs. \ref{Fig:ab_plane} and \ref{Fig:layer_A_B}). Each
of the two cation layers, which are both of the $\theta$-type, are
composed of two different stacks labelled A1 (B1) and A2 (B2) in
layer A (B). The angle between the average planes of cations A1
and A2 is 71.6$^{\circ}$ while the angle between B1 and B2 is
78.0$^{\circ}$. At room temperature, this angle gets an
intermediate value of 73.9$^{\circ}$ \cite{Ly01}. The radical
cation layers also differ in the number of S...S, S...Se and
Se...Se contacts shorter than the sum of the van der Waals radii,
which is larger in layer A than in layer B. In summary, the phase
transition around 240 K corresponds to a degradation of the
crystallographic order. In addition, the solvent molecules can
occupy two different sites, as above mentioned. As a result, the
degree of disorder increases as the temperature is lowered below
the phase transition, as already reported in Ref. \cite{Ly01}.

The calculated electronic band structures and FS's corresponding
to layers A and B are reported in Fig. \ref{Fig:FS}. There are
four donors per repeat unit of the layers so that the band
structures contain four bands mainly built from the HOMO (highest
occupied molecular orbital) of the donors. Since the average
charge of the donors is $+0.5$ these bands should house two holes.
The two  FS's are almost identical but turned approximately by
90$^\circ$. These FS's can be seen as resulting from the
superposition (and hybridization) of a closed loop which, as for
usual $\theta$ phases \cite{Mo99}, may be described either as a
rounded rectangle or as an elongated ellipse. The result is a
Fermi surface containing closed and open portions with very small
hybridization gaps. From the viewpoint of the resulting orbits
network, both of them can be regarded as a linear chain of coupled
orbits, roughly turned around by 90$^\circ$ from each other. This
picture is strongly different from the FS deduced from
calculations at room temperature which yields a network of
compensated electron and holes orbits, as displayed in Fig.
\ref{Fig:FS_roomT}. The area of the closed part of the FS at low
temperature, which correspond to the so called $\alpha$ orbit, is
equal to 22.9$\%$ and 19.6$\%$ of the FBZ area for layers A and B,
respectively. The area of the $\beta$ orbit which can be recovered
by MB is equal to that of the FBZ for both layers.

\subsection{\label{sec:SdH}SdH oscillations}                             

\begin{figure}                                                    
\centering \resizebox{\columnwidth}{!}{\includegraphics*{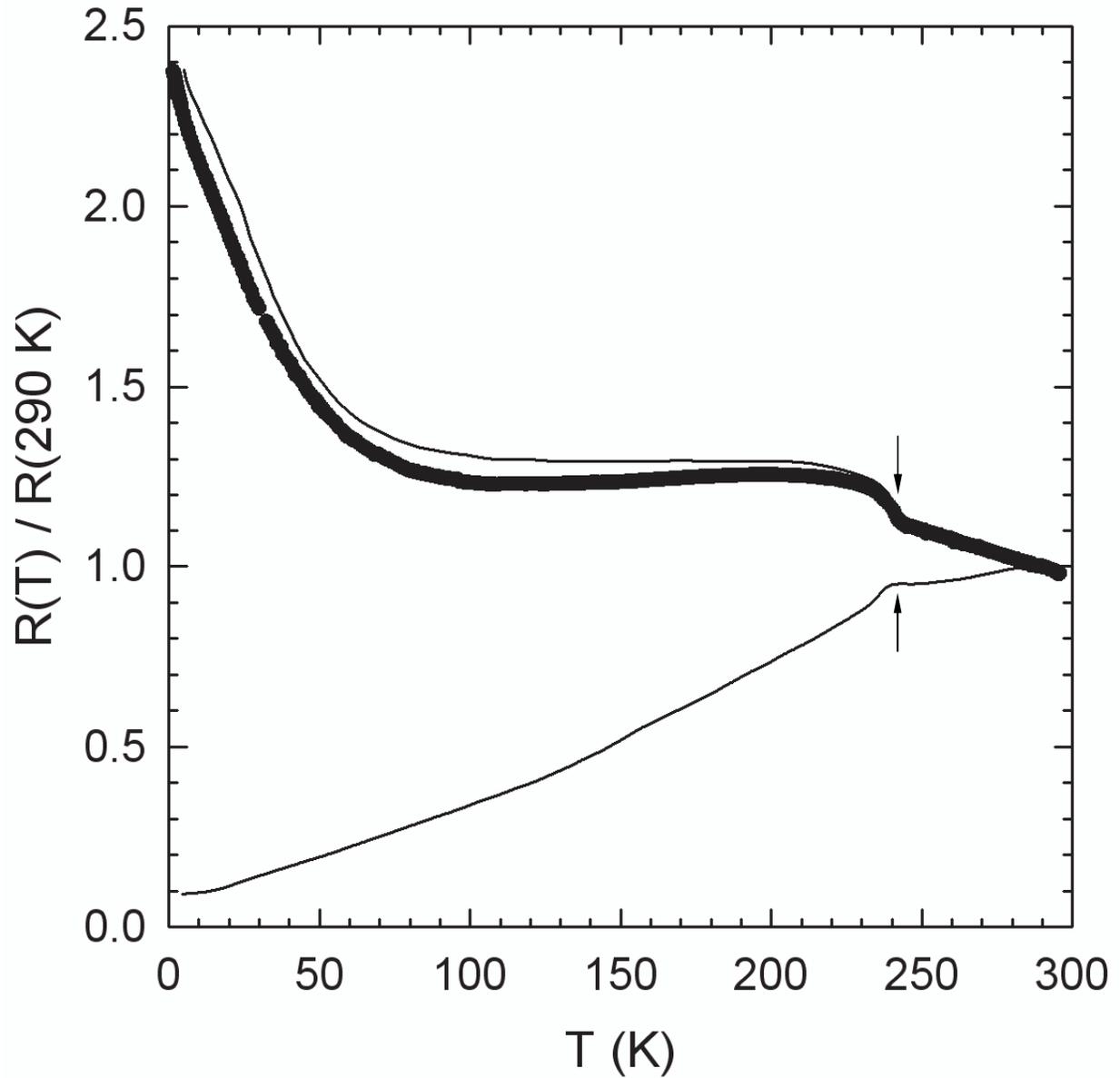}}
\caption{\label{Fig:RT} Temperature dependence of the relative
interlayer resistance of crystal $\# 1$ (symbols). Upper and lower
solid lines are the data for the interlayer and the in-plane
resistances, respectively, from Ref. \cite{Ly01}. The arrows mark
the first order transition (see text).}
\end{figure}

Interlayer zero-field resistance and magnetoresistance data of the
two studied crystals yield consistent results. The temperature
dependence of the zero-field resistance of crystal $\# 1$ is
displayed in Fig. \ref{Fig:RT}. A good agreement with data of Ref.
\cite{Ly01} is observed. Remarkably, the interlayer resistance
exhibits a  behaviour strongly different from that of the in-plane
one. This behaviour has not received any interpretation up to now.
Another salient feature of these data is the kink (marked by
arrows in Fig. \ref{Fig:RT}) which is the signature of the above
discussed phase transition occurring around 240 K.

\begin{table}                                                            
\caption{\label{table}Frequencies, effective masses (in m$_e$
units) and apparent Dingle temperatures linked to the various
Fourier components (index i of Eqs. \ref{LK} to \ref{eq:RS})
appearing in Figure \ref{Fig:TF}.\\}
\begin{tabular}{cccc}
\hline
i &F$_i$ (T)&m$^*_i$&T$^a_{Di}$ (K)\\
\hline 1&37 $\pm$ 1&0.30 $\pm$ 0.05&3.0 $\pm$ 1.2\\
2&81.5 $\pm$ 1.0&0.45 $\pm$ 0.05&11 $\pm$ 3\\
3&198 $\pm$ 4&0.85 $\pm$ 0.10&1.9 $\pm$ 0.8\\
4&545 $\pm$ 2&0.45 $\pm$ 0.10&11 $\pm$ 4\\
5&745 $\pm$ 10&1.15 $\pm$ 0.15&4.5 $\pm$ 1.5\\
6&1270 $\pm$ 30&1.2 $\pm$ 0.2&11 $\pm$ 6\\
7&2900 $\pm$ 100&&\\
8&4200 $\pm$ 200&$\sim$ 2&$\sim$ 20\\
\hline
\end{tabular}
\end{table}

\begin{figure}                                                    
\centering
\resizebox{\columnwidth}{!}{\includegraphics*{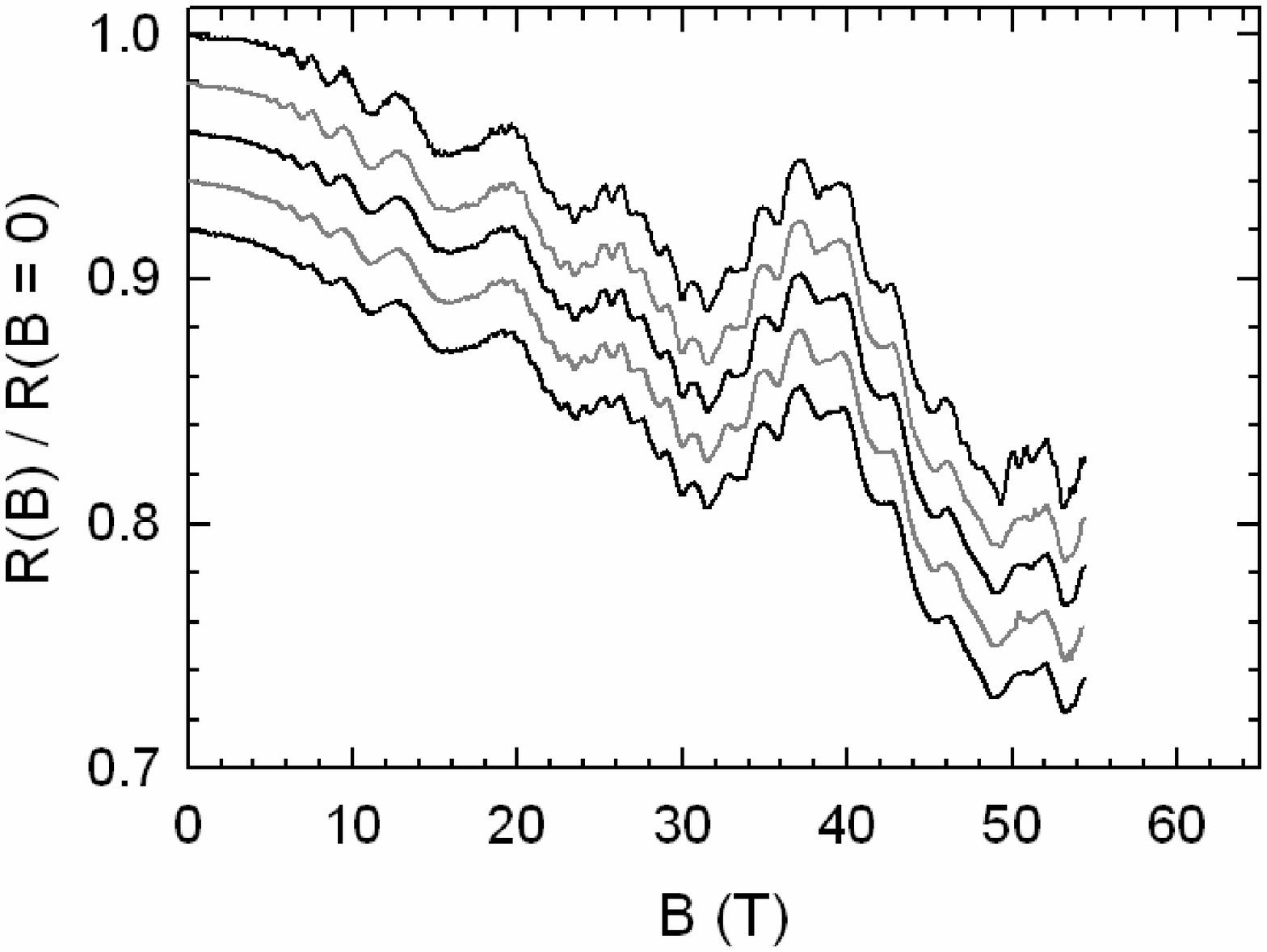}}
\caption{\label{Fig:R(B)} Field-dependent resistance of crystal
$\# 1$ for $\theta$ = 0°. The curves have been shifted down from
each other by 0.02 for clarity.}
\end{figure}

\begin{figure}                                                    
\centering \resizebox{\columnwidth}{!}{\includegraphics*{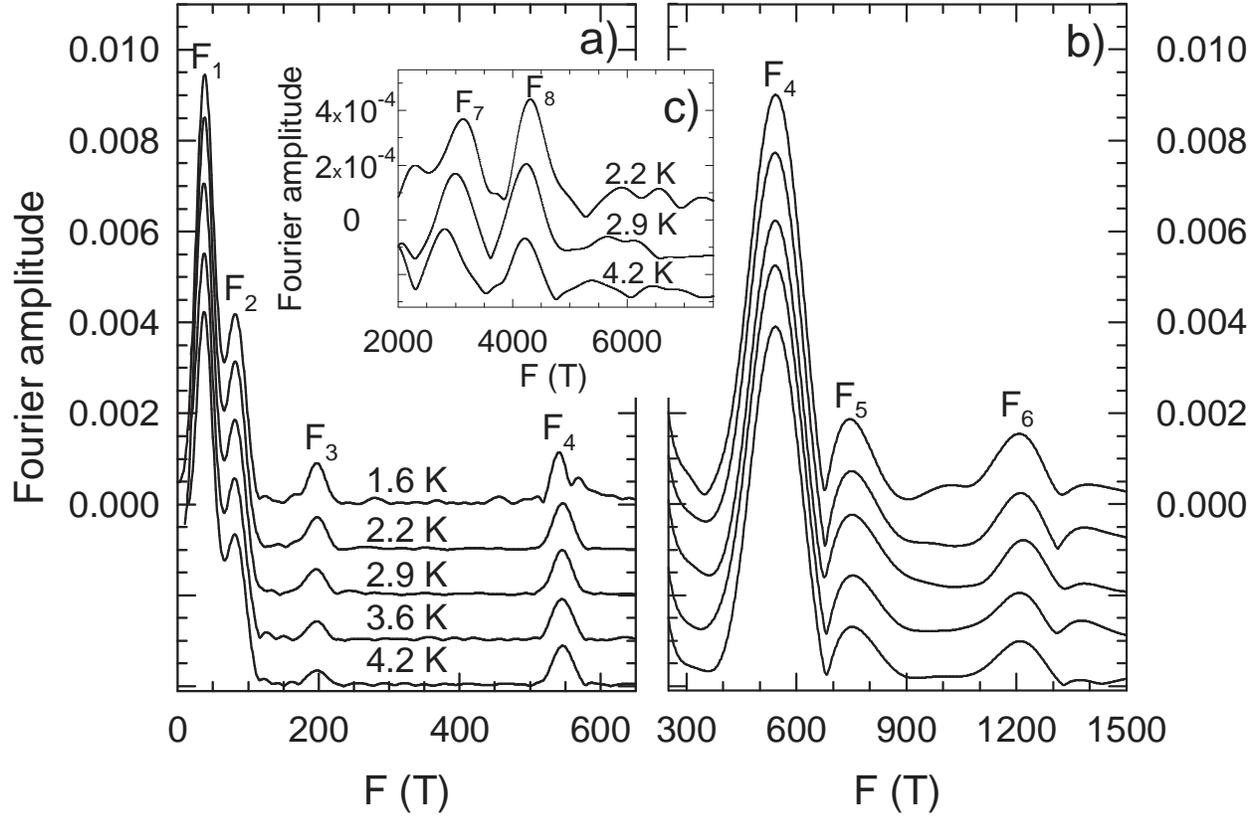}}
\caption{\label{Fig:TF} Fourier spectra of the oscillatory part of
the magnetoresistance data displayed in Fig. \ref{Fig:R(B)}. The
field range is 10 - 54 T, 18 - 54 T and 45 - 54 T in (a), (b) and
(c), respectively. The curves have been shifted down from each
other for clarity.}
\end{figure}

In magnetic field, the interlayer resistance decreases by about 20
$\%$ (see the data of crystal $\# 1$ in Fig. \ref{Fig:R(B)}). This
feature, which was not observed in Ref. \cite{Ly01}, could not be
due to disorder as discussed below. Despite the increase of the
disorder level as the temperature is lowered below the phase
transition (see Section \ref{sec:structures}), quantum
oscillations are clearly resolved. The Fourier spectra of the
oscillatory part of the magnetoresistance is displayed in Fig.
\ref{Fig:TF}. Numerous components can be identified, as it is
observed in many networks of coupled orbits. More precisely, 8
frequencies are observed, of which the values are reported in
Table \ref{table}. It can be noticed that, as already reported in
Ref. \cite{Ly01}, only F$_1$ and F$_3$ can be observed below 15 T.
Roughly speaking, the high frequency oscillations can only be
detected at high field. In particular, the frequencies F$_7$ and
F$_8$, which are in few cases poorly resolved, can only be
observed above about 40 T. Data for crystal $\# 2$, which was
obtained up to 36 T for various directions of the magnetic field,
allow to check that the angle dependence of the frequencies F$_1$
to F$_5$ is consistent with a two-dimensional FS.

\begin{figure}                                                    
\centering \resizebox{\columnwidth}{!}{\includegraphics*{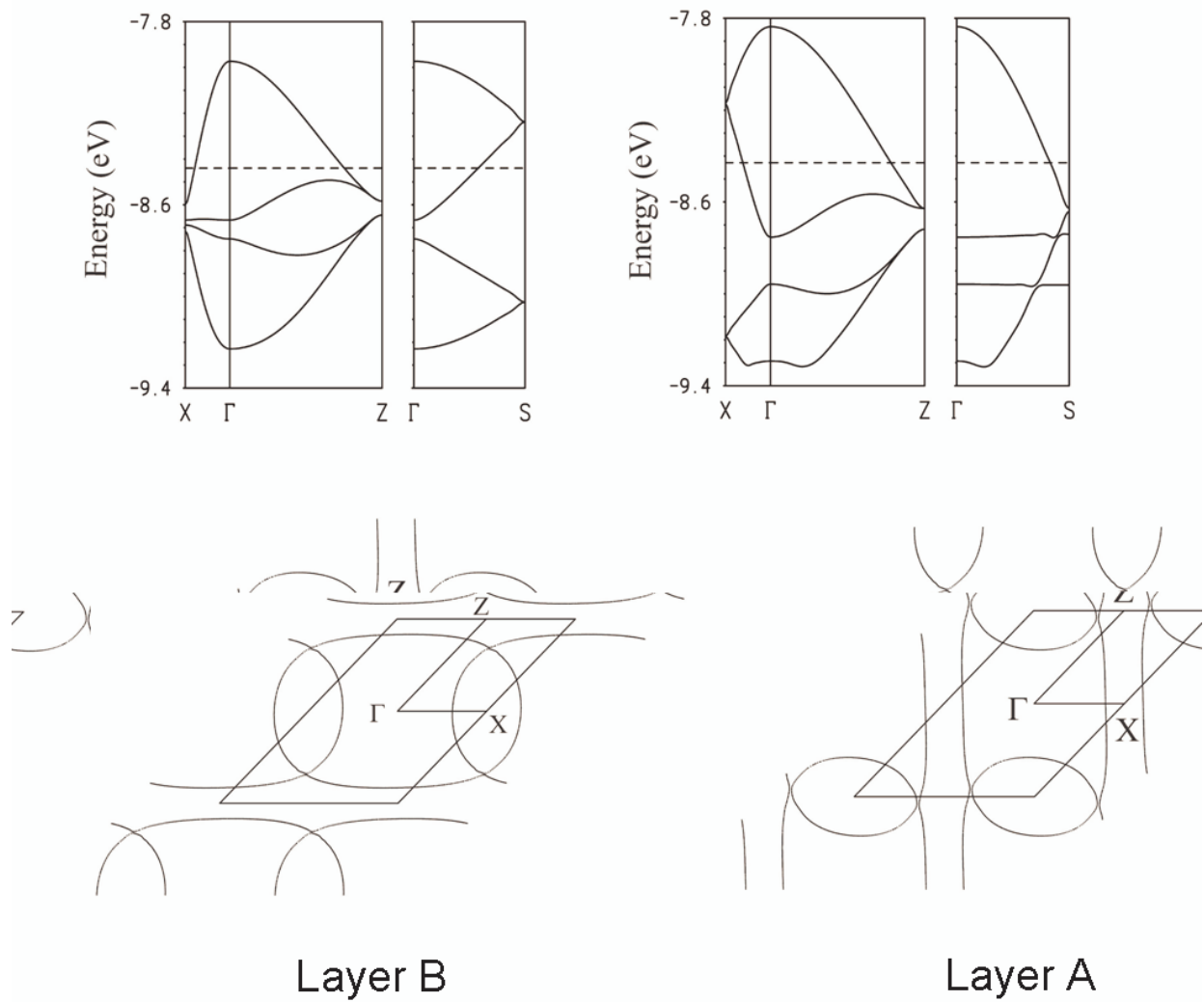}}
\caption{\label{Fig:FS} Electronic band structure, where the
dashed lines refer to the Fermi level, and Fermi surface of the
radical cations layers A and B (see text).}
\end{figure}

As discussed in Ref. \cite{Au04}, whatever the origin of the
frequency combinations liable to be involved in an oscillatory
spectrum (MB-induced closed orbits, QI paths or frequency mixing)
they should be linear combinations of the frequencies linked to
each of the closed orbits and of the $\bigodot$ orbit, namely
F(e), F(h$_1$) F(h$_2$) and F($\bigodot$), in networks of
compensated orbits such as that depicted in Fig.
\ref{Fig:FS_roomT}. In contrast, oscillatory spectra of linear
chains of orbits only involves linear combinations of the
frequencies linked to the closed $\alpha$ and to the MB-induced
$\beta$ orbit, the latter being the analogue of the above
mentioned $\bigodot$ orbit. However, the topology of the FS
displayed in Fig. \ref{Fig:FS} is more complex and the two FS
linked to layer A and B should contribute to the observed
oscillatory spectra.

The only observed frequency that can be attributed to the $\beta$,
or equivalently to the $\bigodot$ orbit, is F$_8$ which
corresponds to an orbit area of 96 $\pm$ 5 $\%$ of the FBZ area at
200 K. In the framework of the band structure calculations at room
temperature, three basic frequencies F(e), F(h$_1$) and F(h$_2$)
with areas of 6, 4 and 2 $\%$ of the FBZ area, respectively,
should be observed, the orbit's compensation yielding F(e) =
F(h$_1$) + F(h$_2$). Such a linear combination is observed in the
present case since F$_3$ + F$_4$ = 743 $\pm$ 6 T which is equal to
F$_5$ within the error bars. In the case where this picture is
relevant, F$_6$ which is equal to F$_4$ + F$_5$ (1290 $\pm$ 12 T)
could be regarded as a frequency combination. However, such
analysis puts aside frequencies F$_1$ and F$_2$. In addition,
F$_5$ corresponds to 17 $\%$ of the FBZ area. This value is much
larger than predicted by the band structure calculations above
mentioned. These features confirm that the FS of the compound
cannot be interpreted on the basis of band structure calculations
at room temperature.

In the framework of the FS calculations of Fig. \ref{Fig:FS}, only
F$_5$ and (or) F$_6$ which correspond to orbits areas of 17 $\%$
and 29 $\%$, respectively, of the FBZ area could be attributed to
the $\alpha$ orbit. In such a case, F$_7$, which is equal to F$_8$
- F$_6$ could correspond to one of the two QI path $\beta -
\alpha$. However, the low frequencies F$_1$ to F$_4$ with areas in
the range from 1 $\%$ to 12 $\%$ cannot be interpreted on the
basis of this FS, unless additional orbits can be considered. For
example, MB-like orbits induced by carriers jump from one layer to
the other could account for frequencies lower than F$_{\alpha}$.
Obviously such an hypothesis requires both experimental and
theoretical confirmation. The possibility of an additional phase
transition at a temperature lower than 200 K, such as density wave
or solvent molecules ordering, could also be considered. For
instance, such a modulation could induce a folding of the FS's
resulting in additional small closed orbits accounting for the
observed discrepancy between the SdH spectra and the present band
structure calculations.



Let us consider now the temperature and field dependence of the
Fourier components reported in Fig. \ref{Fig:TF}. In the framework
of the Lifshitz-Kosevich model \cite{Sh84}, the oscillatory part
of the magnetoresistance of a 2D FS is given by:

\begin{eqnarray}
\label{LK} \frac{R(B)}{R_{background}} = 1 +
\sum_{i}A_{i}cos[2{\pi}(\frac{F_i}{B\cos\theta}-\gamma_i)]
\end{eqnarray}

where $\gamma_i$ is a phase factor. The amplitude of the Fourier
component with frequency F$_i$ is given by A$_{i}$ $\propto$
R$_{Ti}$R$_{Di}$R$_{MBi}$R$_{Si}$. The thermal, Dingle, MB and
spin damping factors are respectively given by:

\begin{eqnarray}
\label{eq:RT}R_{Ti} =
\frac{{\alpha}Tm{_i^*}}{Bcos\theta\sinh[{\alpha}Tm{_i^*}/Bcos\theta
]} \\\label{eq:RD}R_{Di} =
exp[-{\alpha}T_{Di}m{_i^*}/Bcos\theta]\\
\label{eq:RMB} R_{MBi} =
exp(-\frac{t_iB_{0}}{2B})[1-exp(-\frac{B_{0}}{B})]^{b_i/2}\\
\label{eq:RS} R_{Si} = \mid\cos(\pi \mu_i/\cos\theta)\mid
\end{eqnarray}

where $\alpha$ = 2$\pi^2$m$_e$k$_B$/e$\hbar$ ($\simeq$ 14.69 T/K),
$\mu_i$ = g$^*$m${_i^*}$/2, g$^*$ is the effective Land\'{e}
factor, m$_i^*$ is the effective mass normalized to the free
electron mass m$_e$, T$_{Di}$ is the Dingle temperature and
B$_{0}$ is the MB field which is assumed here to be the same for
all the MB junctions (this point is discussed in Ref.
\cite{Vi03}). Integers $t_i$ and $b_i$ are respectively the number
of tunnelling and Bragg reflections encountered along
the path of the quasiparticle.   \\

\begin{figure}                                                    
\centering
\resizebox{\columnwidth}{!}{\includegraphics*{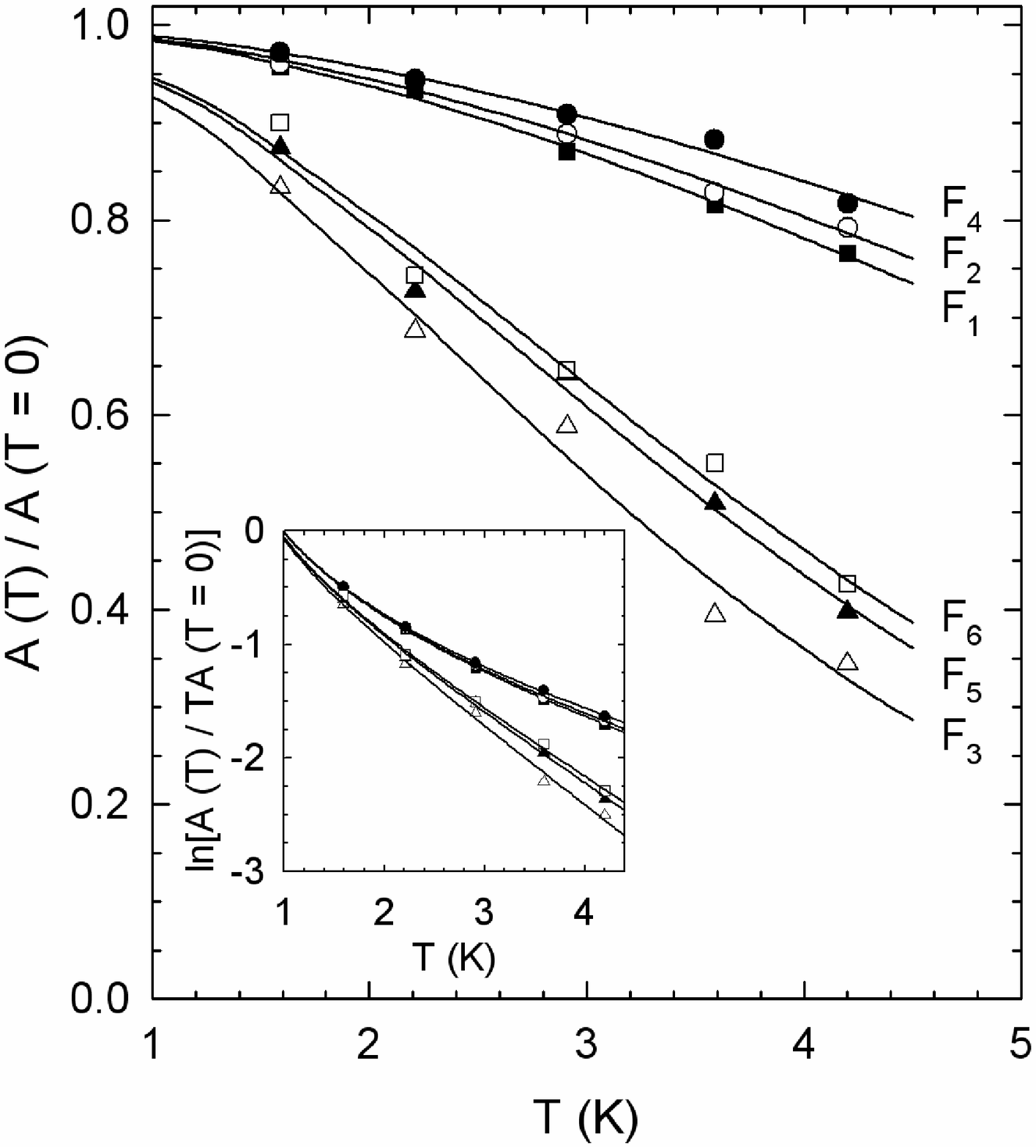}}
\caption{\label{Fig:A(T)} Temperature dependence of the Fourier
components' amplitude. The inset displays the data in a
semi-logarithmic scale. The mean field value is 14 T, 20 T, 20 T,
25 T, 28 T and 30 T for F$_1$ to F$_6$, respectively. Solid lines
are best fits of Eq. \ref{LK} to the data.}
\end{figure}

\begin{figure}                                                    
\centering
\resizebox{\columnwidth}{!}{\includegraphics*{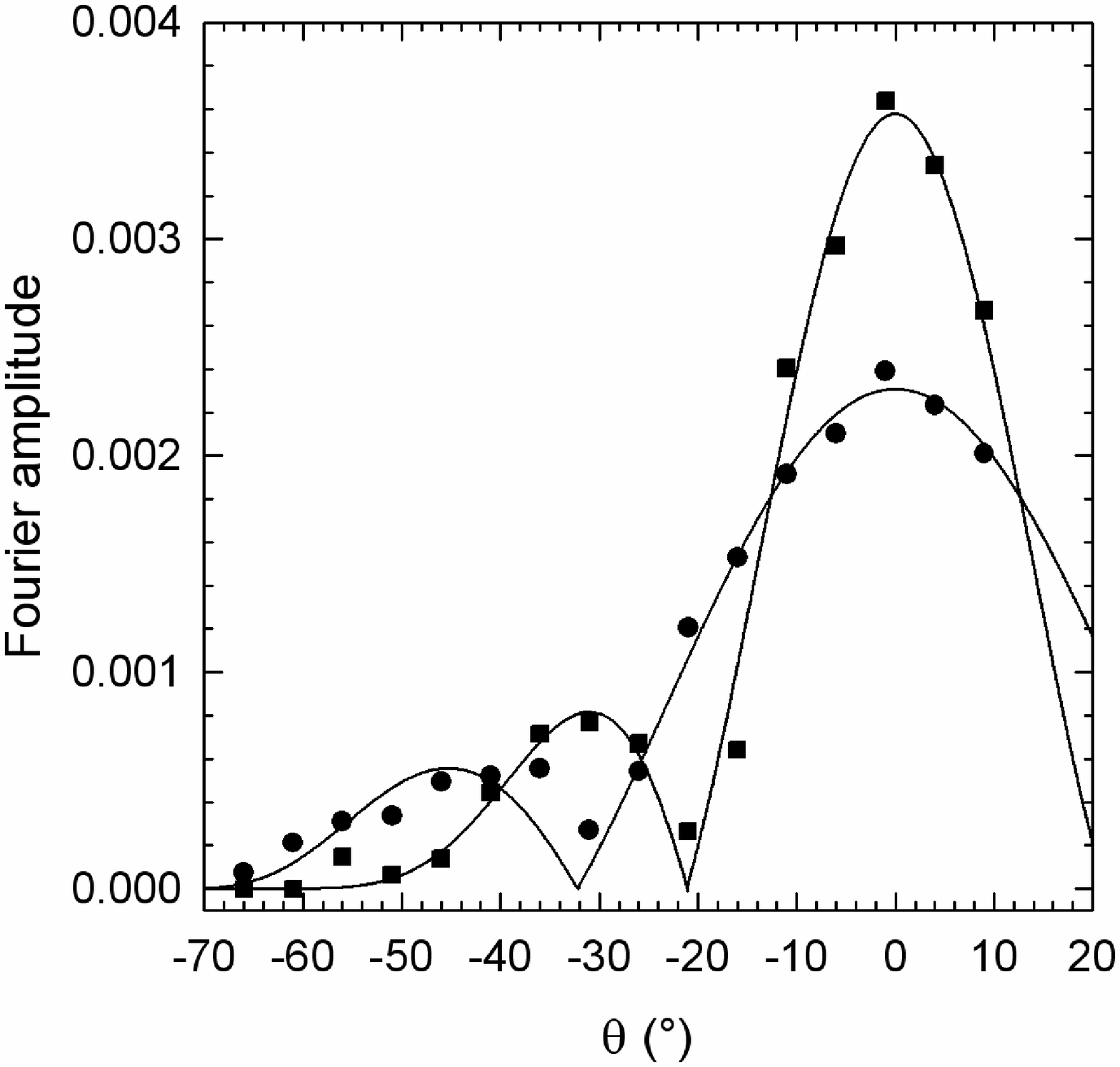}}
\caption{\label{Fig:spin0} Angle dependence of the amplitude of
the Fourier components F$_2$ (circles) and F$_4$ (squares) for
crystal $\#2$. The mean magnetic field value is 16.8 and 26.6 T,
for F$_2$ and F$_4$, respectively. Solid lines are best fits of
Eq. \ref{LK} to the data obtained with $\mu$ = 0.42 and 0.47, for
F$_2$ and F$_4$, respectively (see text).}
\end{figure}

As displayed in Fig. \ref{Fig:A(T)}, the temperature dependence of
the various Fourier components observed in Fig. \ref{Fig:TF} is in
agreement with the LK formalism. The deduced effective masses are
reported in Table \ref{table}. It has been checked that they
remain field-independent within the error bars in the range where
the Fourier components are detectable. Effective masses can also
be derived from the angle dependence of the oscillation amplitude
which involves the spin damping factor (see Eq. \ref{eq:RS}).
Examples are given in Fig. \ref{Fig:spin0}: a good agreement is
obtained yielding $\mu_i$ values that are equal to m$^*_i$ within
the error bars. This result indicates that g$^*$ is close to 2
which is in line with a Fermi liquid picture. The values of the
effective mass for F$_1$ to F$_6$ are in the range from 0.3 to
about 1.2 m$_e$. Similar rather low values are also observed in
various q-2D organic metals for which the frequencies involved in
the SdH spectra are of the same order of magnitude. This is, in
particular, the case of the family
$\beta$''-(BEDT-TTF)$_4$(NH$_4$)[M(C$_2$O$_4$)$_3$]$\cdot$DMF (M =
Fe, Cr) of which the FS is rather complex \cite{Au04,Vi06} as it
is in the present case.

\begin{figure}                                                    
\centering
\resizebox{\columnwidth}{!}{\includegraphics*{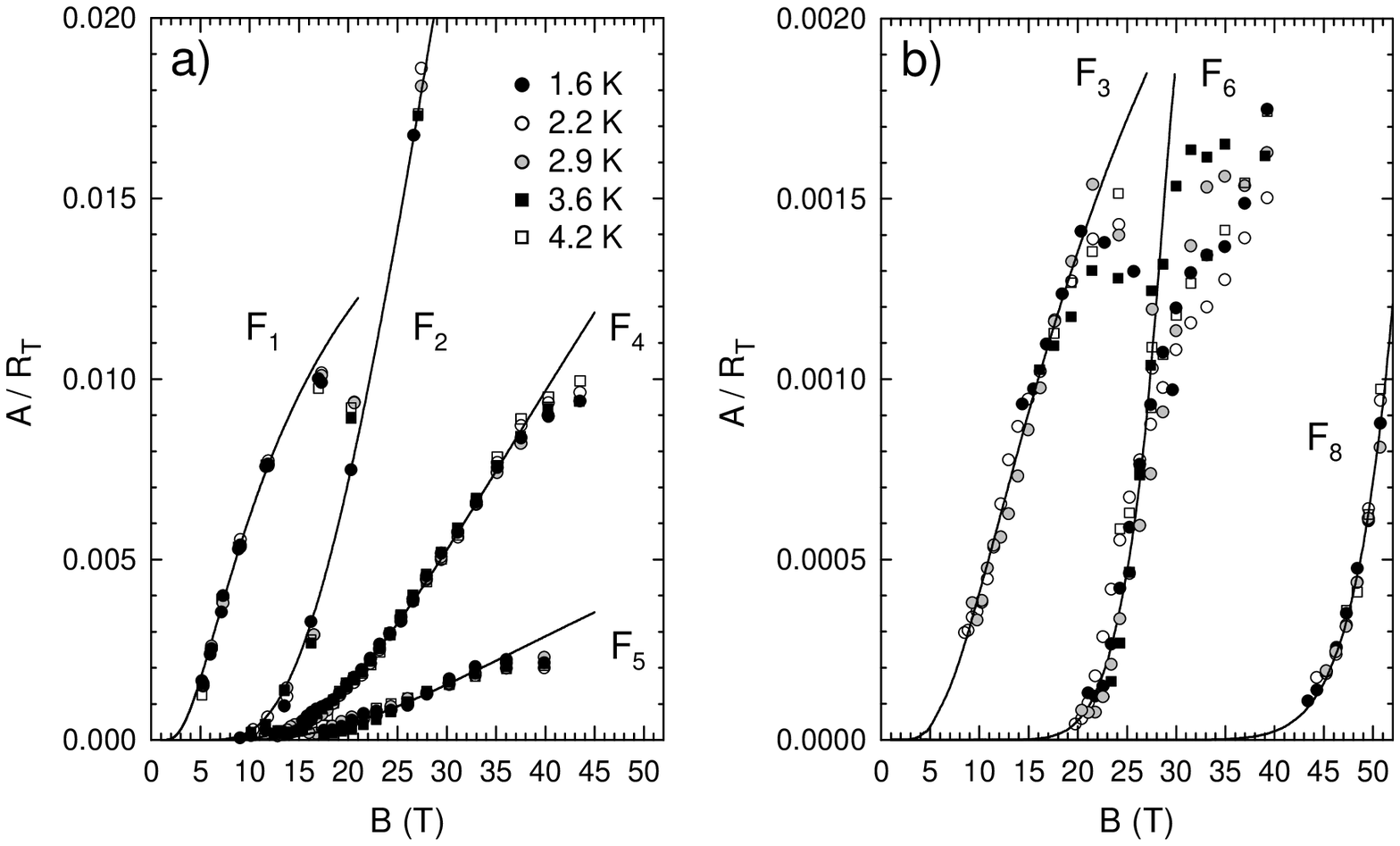}}
\caption{\label{Fig:Dingle} Field dependence of the Fourier's
components amplitude normalized to the Fermi-Dirac smearing
damping factor (R$_T$). Solid lines are best fits of Eq.
\ref{Eq:AsurRT} to the data assuming b$_i$ = 0 (see text).}
\end{figure}

Additional information on the oscillatory spectrum can be obtained
through the field dependence of the oscillations amplitude. For
$\theta$ = 0, Eq. $\ref{LK}$ can be rewritten as:

\begin{eqnarray}                                                       
\label{Eq:AsurRT}\frac{A_i}{R_{Ti}} \propto
exp[-\frac{{\alpha}T_{Di}m{_i^*} + t_iB_{0}/2}{B}] \nonumber\\
\times [1-exp(-\frac{B_{0}}{B})]^{\frac{b_i}{2}}
\end{eqnarray}

which yields the temperature-independent part of the amplitude,
provided the relevant effective mass has been determined. At low
magnetic field or in the case where b$_i$ = 0, Eq. \ref{Eq:AsurRT}
simplifies as A$_i$ / R$_{Ti} \propto$
exp[-($\alpha$T$_{Di}$m$_i^*$ + t$_i$B$_{0}$/2)/B]. In the case
where t$_i$ = 0, i. e. for a basic orbit, the field dependence of
A$_i$ / R$_{Ti}$ yields the Dingle temperature, as usual. In
contrast, in the case where t$_i$ $\neq$ 0, only an apparent
Dingle temperature, higher than T$_{Di}$, can be derived
(T$_{Di}^a$ = T$_{Di}$ + t$_i$B$_{0}$/2$\alpha$m$_i^*$). The field
dependence of A$_i$ / R$_{Ti}$ is displayed in Fig.
\ref{Fig:Dingle} for various Fourier components. It can be
remarked first that, in agreement with the predictions of Eq.
\ref{Eq:AsurRT}, this parameter is actually
temperature-independent, at least in the low field range. Solid
lines in this figure are best fits to the data assuming b$_i$ = 0
which actually holds in the low field range as above discussed.
The deduced values of T$^a_{Di}$ (see Table \ref{table}) are very
disparate. It should be noticed that the set of values obtained
for crystal $\#2$ is in good agreement with the data of Table
\ref{table}. Furthermore, a re-analysis of the low field data of
Ref. \cite{Ly01} yields T$_{D1}^a \simeq$ 4 K which is close to,
and in any case not lower than, the present data. This result
suggests that the negative magnetoresistance observed in Fig
$\ref{Fig:R(B)}$ could not be due to disorder. Downward deviations
of the data from the fittings of Fig. \ref{Fig:Dingle}, that could
be due to Bragg reflections (in which case b$_i \neq$ 0) are
observed at high field for some of the Fourier components. The
very high values of T$_{Di}^a$ observed for F$_2$, F$_4$, F$_6$
and F$_8$ could indeed be the signature of MB. This should be in
particular the case of F$_8$ in the case where it actually
corresponds to the large orbit from which the FS is built. In line
with this assumption, it is tempting to consider that F$_1$, F$_3$
and F$_5$ correspond to basic orbits. However, the field
dependence of the amplitude of Fourier components corresponding to
MB orbits can be strongly influenced by the frequency mixing
phenomenon \cite{Pr02,Vi03,Au05} which hampers any definite
conclusion in the absence of reliable FS determination at low
temperature. Nevertheless, it is plausible that some of the
observed Fourier components correspond to MB and (or) frequency
mixing. In addition, the occurrence of a new type of MB-like
quantum oscillation linked to the considered non-trivial FS cannot
be excluded, although this latter hypothesis requires both
experimental and theoretical confirmation.
\\

\section{\label{sec:Conclusion}Summary and Conclusion}

The crystal and electronic band structures of the organic metal
$\theta$-(BETS)$_4$HgBr$_4$(C$_6$H$_5$Cl) have been determined at
a temperature of 200 K, i. e.  below the phase transition which
occurs at about 240 K. X-ray data indicate a modification of the
crystal structure the symmetry of which is lowered from quadratic
to monoclinic. Strikingly, two different types of cation layers
which alternate in the direction perpendicular to the conducting
plane are observed. The resulting non-trivial Fermi surface (FS),
which can be regarded as a network of compensated electron and
hole orbits according to band structure calculations at room
temperature, turns to a set of two alternating linear chain of
orbits at low temperature.

The SdH oscillations spectra measured up to 54 T reveal eight
Fourier components. Their temperature and field (both magnitude
and orientation) dependence are in good agreement with the LK
formalism. Owing to their field dependence, it is likely that some
of the observed Fourier components correspond to either MB orbits
or frequency mixing. In any case, the oscillatory data suggest a
complex FS that is not in agreement with band structure
calculations at room temperature. Band structure calculations at
200 K cannot fully account for the data either, unless additional
MB-like orbits induced by carriers jump from one layer to the
other can be considered. An additional phase transition below 200
K might also be considered to be at the origin of the
disagreement. More work at low temperature seems to be mandatory
to fully understand the intriguing SdH oscillations spectra of the
q-2D organic metal $\theta$-(BETS)$_4$HgBr$_4$(C$_6$H$_5$Cl).

\begin{acknowledgments}
This work was supported by Euromagnet under the European Union
contract R113-CT-2004-506239, the CNRS-RAS cooperation agreement
$\#$ 16390, the CNRS-CSIC exchange program (2005FR0019), MEC-Spain
(FIS2006- 12117-C04-01) and Generalitat de Catalunya (Project 2005
SGR 683). Part of the computations described in this work were
carried out using the resources of CESCA and CEPBA. RBL thanks
RFBR 07-02-00311 for support.
\end{acknowledgments}

\newpage

\end{document}